\documentclass[12pt]{article}
\usepackage{graphicx}
\usepackage{amssymb,amsmath}

\usepackage{epsfig,cite}
\newcommand{\be}{\begin{equation}}
\newcommand{\ee}{\end{equation}}
\newcommand{\bea}{\begin{eqnarray}}
\newcommand{\eea}{\end{eqnarray}}

\newcommand{\mo}{\mathcal{O}}

\newcommand{\lc}{\left[}
\newcommand{\rc}{\right]}
\newcommand{\lp}{\left(}
\newcommand{\rp}{\right)}
\newcommand{\bc}{\begin{center}}
\newcommand{\ec}{\end{center}}
\def\epm#1#2{\hbox{${\lower1pt\hbox{$\scriptstyle +~#1$}}
\atop {\raise1pt\hbox{$\scriptstyle -~#2$}}$}}

\newcommand{\J}{\mathcal{J}}
\newcommand{\E}{\mathcal{E}}

\def\a{\alpha}
\def\b{\beta}

\def\s{\sigma}
\def\w{\omega}

\def\appendix#1{
  \addtocounter{section}{1}
  \setcounter{equation}{0}
  \renewcommand{\thesection}{\Alph{section}}
  \section*{Appendix \thesection\protect\indent \parbox[t]{11.15cm}
  {#1} }
  \addcontentsline{toc}{section}{Appendix \thesection\ \ \ #1}
  }

\begin{document}

 
\begin{flushright}
 
{\tt hep-th/0408174} 
\\  UB--ECM--PF 04/20

\end{flushright}

\begin{center}

\vspace*{.6cm}

 {\bf{\Large Non-perturbative  states in type II superstring theory \\ 
from classical spinning membranes}}

\vspace*{1.3cm}

{\bf  Jan Brugues$^1$, Joan Rojo$^1$ and Jorge G. Russo$^{1,2}$,}

\vspace{0.5cm}

~$^1$ Departament d'Estructura i Constituents de la Mat\`eria, \\
Universitat de Barcelona, Diagonal 647, E-08028 Barcelona, Spain

\medskip

$^2$ Instituci\' o Catalana de Recerca i Estudis
  Avan\c{c}ats (ICREA)

\vspace*{2.5cm}
                                                                 
{\bf Abstract}

\end{center}
\noindent

We find a new family of exact solutions in membrane  theory, representing
 toroidal membranes spinning in several planes. They have energy square
proportional to the sum of the different angular momenta, generalizing
Regge-type string solutions to membrane theory.
By compactifying the eleven dimensional theory on a circle and on a
torus, we identify a family of new non-perturbative
states of type IIA and type IIB superstring theory
 (which contains
the perturbative spinning string solutions of type II string theory 
as a particular case). The solution represents a 
spinning bound state of D branes and fundamental strings.
Then we find similar solutions for membranes on $AdS_7\times S^4$
and $AdS_4\times S^7$.
We also consider the analogous solutions in $SU(N)$ matrix
theory, and compute the energy. They can be interpreted as
rotating open strings with D0 branes attached to their endpoints.

\vfill 
\begin{flushleft} 
July 2004 
\end{flushleft}

\eject   
\tableofcontents

\section{Introduction}

Understanding new aspects of M-theory may eventually lead to
a powerful setup to uncover and clarify
the non-perturbative physics of string theory
(for a review of M-theory, see \cite{mtheory}).
An important piece of information about the theory is its mass spectrum.
In eleven uncompact dimensions, 
determining the spectrum of quantum states is a hard problem, there is
no coupling constant and
gravitational effects are always important.
In compactifying the eleven dimensional coordinate $X^{10}$ on a
circle of radius $R_{10}$, one makes
contact with type IIA superstring theory with coupling
$g_{\rm IIA}=2\pi R_{10}^3/l_{P}^3$, where $l_{P}$ is 
the eleven-dimensional Planck length \cite{town2,witten}.
For small radius, M-theory describes a weakly-coupled string theory
and it is possible to compute masses without the complication
of gravitational interactions, provided they are of order 
$M=O(g_{\rm IIA}^{-s})$, $s<2 $ so that the gravitational forces
(proportional to $ g_{\rm IIA}^2 M$)
are negligible  as $g_{\rm IIA}\to 0$. In particular, this is the reason
why there exists a simple weak coupling description for  D0 branes, which
have masses $O(g_{\rm IIA}^{-1})$.

M-theory is known to contain membranes and five branes (see e.g. 
\cite{mtheory,duff}).
In this paper we will find new classical solutions in supermembrane theory
\cite{townsend,townsend2}
and compute their energy. 
In flat space, these solutions are similar to the spinning string solutions of
\cite{CIR}, which were found to be highly stable quantum mechanically
in 
weakly-coupled string theory. For a special choice of  quantum
numbers
(such that the solution depends only on one spatial membrane
coordinate)
the solution reduces to the string solution of \cite{CIR}.
For general quantum numbers, the energy has
non-trivial dependence on the coupling constant $g_{\rm IIA}$ of the
form
$E\sim O(g_{\rm IIA}^{-1})$. This means that for small radius $R_{10}$
one can ignore gravitational effects, so  these solutions describe
genuine quantum states with high-excitation quantum numbers in
M-theory,
and new non-perturbative states in type IIA string theory.
By further compactifying the coordinate $X^9$ on  a circle
of radius $R_9$, one makes contact with type IIB string theory.
Ten-dimensional type IIB theory arises as  M-theory
on the torus $(X^9,X^{10})$ in the limit that
the area $R_9R_{10}$ goes to zero at fixed $g_{\rm IIB}={R_{10}\over
  R_9}$. Membranes states can lead to various types of
non-perturbative objects of type IIB theory, such as
bound  states of fundamental strings
and D strings \cite{schwarz,waves,bps}.
The spinning solutions presented here lead to  a new class of
non-perturbative states of type II string theory, representing
rotating objects which do not have a pure fundamental string
interpretation. They have in general D brane charges and fundamental
string charges.

We then consider similar spinning solutions for membrane theory
in $AdS_7\times S^4$ and in  $AdS_4\times S^7$.
These solutions are the membrane analogue of the
general class of spinning string solutions found
in $AdS_5\times S^5$  in \cite{ART} (which includes 
some spinning string solutions 
appeared earlier in \cite{FT,AFRT}). 
The energy admits a simple expansion at large angular momenta.

It has been proposed \cite{mtheoryconj} that M-theory in the
light-cone frame  is described by Matrix theory, which can be viewed as a
regularized theory of the supersymmetric membrane \cite{WHN} 
 (for reviews see
\cite{susskind,washi}). 
The  fundamental degrees
of freedom can be viewed as the D0 branes, which in the light-cone frame
are expected to capture all the complicated dynamics of M theory.
The Lagrangian is that of  supersymmetric quantum mechanics described by
0+1 super Yang-Mills theory with sixteen supersymmetries.
In the last part of the paper we consider  ``spinning'' solutions
in Matrix theory, found in  \cite{AFP}, which are analogous to the
spinning membrane solutions. 
Here they are 
slightly generalized to incorporate
rotation in four different planes, and in addition we compute
the energy, which was not done in  \cite{AFP}.

There is an extensive literature on classical solutions in membrane/matrix
theory. The reader can look at \cite{townsend2,HopNic} and more recently
\cite{hoppe2,AFP,ali1,ali2,nunez,bozh,hoppe}. 
Some of the present solutions are similar to the ones obtained in
\cite{AFP}.
 In addition, we present generalizations and many new solutions 
that include momentum  and winding in the
directions $X_9$, $X_{10}$, which allows to make contact with type II
string theories. These are in fact the most interesting solutions that
give rise to new non-perturbative states with rotation 
in type II string theory.

  \setcounter{equation}{0}
\section{Spinning membranes solutions in flat space}

\subsection{General rotating ansatz}

The starting point is the  bosonic part of the action for the
supermembrane \cite{townsend} in flat space
\be
\label{action}
S=-\frac{T_2}{2}
\int d^3\xi \lp \sqrt{-h}h^{\alpha\beta}\partial_{\alpha}X^{\mu}
\partial_{\beta}X^{\nu}\eta_{\mu\nu} -\sqrt{-h}\rp \ , \quad
\mu=0,\ldots,10
\ .
\ee
$T_2$ is the membrane tension, $T_2=(2\pi l_P^3)^{-1}$.
The equations of motion for
the metric on the membrane worldvolume gives
\be
h_{\alpha\beta}=\frac{\partial X^{\mu}}{\partial \xi^\alpha}
\frac{\partial X^{\nu}}{\partial\xi^\beta}\eta_{\mu\nu}, \qquad
\alpha,\beta=0,1,2\ .
\ee
In what follows we use the notation $(\xi^0,\xi^1,\xi^2)\equiv (\tau
,\sigma,\rho )$. Treating the $X$ and $h$ as independent fields, we 
now fix the gauge in the usual way \cite{townsend2,bars}, by setting
\be\label{hcon}
h_{\tau\sigma}=h_{\tau\rho}=0\ ,\quad h_{\tau\tau}L^2=-g\ ,
\ee
where $L$ is an arbitrary constant with units of length and
$g$ the determinant of the spacelike part of the induced metric. 
The action (\ref{action})  becomes
\be\label{act2}
S=-\frac{T_2}{2}\int d^3\xi \lp 
-L \partial_{\tau}X^{\mu}\partial_{\tau}X^{\nu}\eta_{\mu\nu}+
\frac{g}{L}\lp h^{ij}\partial_{i}X^{\mu}\partial_{j}X^{\nu}\eta_{\mu\nu} 
-1\rp\rp \ .
\ee
Using the equation of motion for the induced metric, 
$h_{ij}=\partial_{i}X^{\mu}\partial_{j}X^{\nu}\eta_{\mu\nu}$
the action (\ref{act2}) 
reads
\be\label{actgen}
S=-\frac{T_2}{2}\int d^3\xi \lp -L \partial_{\tau}X^{\mu}\partial_{\tau}X_{\mu}
+\frac{1}{2L}\{X_{\mu},X_{\nu}\}\{X^{\mu},X^{\nu}\}\rp \ ,
\ee
where the Poisson bracket is defined as usual as
\be
\{f,g\}\equiv \epsilon^{rs}\partial_rf\partial_s g\ ,\ \ \ \ 
r,s=\s ,\rho\ ,
\ee
for any two differentiable functions $f,g$ on the two dimensional
manifold.

In terms of the $X^\mu $ fields, the constraints for the induced metric (\ref{hcon})
that must be imposed to the solutions are
\begin{align}
\label{memcos1}
&\eta_{\mu\nu}\frac{\partial X^{\mu}}{\partial \tau}
\frac{\partial X^{\nu}}{\partial \sigma}=
\eta_{\mu\nu}\frac{\partial X^{\mu}}{\partial \tau}
\frac{\partial X^{\nu}}{\partial \rho}=0\ ,
\\
\label{memcos2}
&L^2 \eta_{\mu\nu}\frac{\partial X^{\mu}}{\partial \tau}
\frac{\partial X^{\nu}}{\partial \tau}=\lp
\eta_{\mu\nu}\frac{\partial X^{\mu}}{\partial \sigma}
\frac{\partial X^{\nu}}{\partial \rho}\rp^2-
\lp  \eta_{\mu\nu}\frac{\partial X^{\mu}}{\partial \sigma}
\frac{\partial X^{\nu}}{\partial \sigma} \rp
\lp  \eta_{\mu\nu}\frac{\partial X^{\mu}}{\partial \rho}
\frac{\partial X^{\nu}}{\partial \rho} \rp\ .
\end{align}

Now we consider a general rotating ansatz of the following form
\bea
X_0 &=& \kappa \tau \ ,
\nonumber \\
Z_1(\tau,\sigma,\rho)&=&X_1+iX_2=r_1(\sigma,\rho)e^{i\omega_1\tau+
i\alpha_1(\sigma,\rho)}
\nonumber  \ ,\\
Z_2(\tau,\sigma,\rho)&=&X_3+iX_4=r_2(\sigma,\rho)e^{i\omega_2\tau+
i\alpha_2(\sigma,\rho)}
\nonumber  \ ,\\
Z_3(\tau,\sigma,\rho)&=&X_5+iX_6=r_3(\sigma,\rho)e^{i\omega_3\tau+
i\alpha_3(\sigma,\rho)}
\nonumber  \ ,  \\
Z_4(\tau,\sigma,\rho)&=&X_7+iX_8=r_4(\sigma,\rho)e^{i\omega_4\tau+
i\alpha_4(\sigma,\rho)}
 \label{ansatz} \ , \\
X_9 &=& X_{10} =0 \nonumber \ .  
\eea
This represents a membrane  spinning in four orthogonal planes
$Z_1,\ldots,Z_4$. 
Using this ansatz,  the action for the membrane (\ref{actgen}) reads
\be
S=-\frac{T_2}{2}
\int d^3\xi \Bigg( L(\kappa^2-\sum_{a=1}^4\omega_a^2r_a^2)+\frac{1}{2L}
\sum_{a,b=1}^4
\lp \{r_a,r_b \}^{2}+
r_a^2r_b^2\{\alpha_a,\alpha_b  \}^{2}
+2r_a^2\{\alpha_a,r_b \}^2\rp\Bigg)\ .
\ee
The equations of motion for  the radial
and the angular coordinates are given by
\begin{align}
  &\begin{aligned}\label{eomrad}
    r_{c}\left(L^{2} \w_{c}^{2}-\sum_{a=1}^4(r_{a}^{2}\{\a_{c},\a_{a}\}^{2}
    +\{r_{a},\a_{c}\}^{2})\right)
    +\sum_{a=1}^4\left(\{\{r_{c},r_{a}\},r_{a}\}
    +r_{a}^{2}\{\{r_{c},\a_{a}\},\a_{a}\}\right)=0
  \end{aligned}
  \\
  &\begin{aligned}\label{eomang}
    \sum_{a=1}^4\left( r_{a}^{2}\{\{\a_{c},\a_{a}\},\a_{a}\}
    +\{\{\a_{c},r_{a}\},r_{a}\}\right)=0\ .
   \end{aligned}
\end{align}
Inserting the ansatz (\ref{ansatz}), 
 the constraints (\ref{memcos1}),  (\ref{memcos2})  take the form
\be
\sum_{a=1}^4\omega_ar_a^2\partial_{\sigma}\alpha_a=
\sum_{a=1}^4\omega_ar_a^2\partial_{\rho}\alpha_a=0\ ,
\ee
\be
L^{2}(\kappa^2-\sum_{a=1}^4\omega_a^2r_a^2)=
\sum_{a,b=1}^4
\lp \frac{1}{2}\{r_a,r_b \}^{2}+
\frac{1}{2}r_a^2r_b^2\{\alpha_a,\alpha_b  \}^{2}
+r_a^2\{\alpha_a,r_b \}^2\rp\ .
\ee

\subsection{Constant radius solution}

A simplification of our original ansatz (that generalizes
the rotating string solutions of  \cite{CIR} to membrane theory)
is considering solutions of constant radius and with the phases
$\alpha_a $
depending linearly on $\sigma, \ \rho$. Assuming that the $r_a$ are
constants,
the equation of motion (\ref{eomang}) for the phases
becomes
\begin{align}
  \sum_{a=1}^4r_{a}^{2}\{\{\a_{b},\a_{a}\},\a_{a}\}=0\ .
\end{align}
This is indeed solved by $\alpha_a(\sigma,\rho)=k_a\sigma+l_a\rho$, i.e.,

\be \label{Za}
Z_{a}(\tau,\sigma,\rho)=r_{a}e^{i\w_{a}\tau+i(k_a\sigma + l_a\rho)}\
,
\quad a=1,\ldots,4 \ ,
\ee
with $k_a, \ l_a$ integer numbers,
as implied by the periodicity condition  satisfied by
a closed membrane.
The equation of motion (\ref{eomrad}) of the radial
coordinates leads to
\be \label{consomega}
L\w_b^2-\frac{1}{L}\sum_{a=1}^4r_a^2(l_bk_a-l_ak_b)^2=0\ .
\ee
This equation determines the frequencies $\omega_b$ 
as a function of the radii $r_a$ and
the different winding numbers. 

Using the ansatz (\ref{Za}), 
the  constraints can be written as
\be
\label{consans1}
\sum_{a=1}^4\omega_ar_a^2k_a=\sum_{a=1}^4\omega_ar_a^2l_a
=0\ ,
\ee
\be\label{consans2}
  L^{2}(\kappa^2-\sum_{a=1}^4\omega_a^2r_a^2)=\sum_{a,b=1}^4
r_a^2r_b^2(k_{a}l_{b}-k_{b}l_{a})^{2}\ .
\ee
Combining with eq. (\ref{consomega}), we get
\be
\kappa^{2}=2\sum_{a=1}^{4} r_{a}^{2}\w_{a}^{2} \ .
\label{cass}
\ee

The solution (\ref{Za}) appeared in \cite{AFP} in the case of 
light-cone gauge membrane theory.
In the light-cone gauge 
the constraints (\ref{consans1}) (which lead to
 constraints on the angular momenta and winding numbers, see
 (\ref{cffg})) 
is absent.
The reason is the following. This constraint comes from
(\ref{memcos1}).
In the light cone gauge $X^+=p_+\tau $, this gives $p_+ \partial _s
X^-=\partial_\tau X^i \partial_s X^i$, which can be solved for $X^-$
provided the integrability condition $\{ \dot X^i, X^i\} =0$ is
satisfied, which is automatically the case for the solution (\ref{Za}).
In the present case, instead, 
$X_0=\kappa \tau $, $X_9=X_{10}=0$, and this
physical membrane state (\ref{Za}) 
exists only if the angular momenta and winding numbers are constrained by
 (\ref{consans1}) (i.e. (\ref{cffg})). 
The light-cone gauge represents an extreme physical 
situation of
infinite momentum, where the theory typically gets simplified.
In section 2.4 we shall consider a
generalization
where the constraint (\ref{memcos1}) can be solved by adding winding
and linear momentum in the directions $X_9,\ X_{10}$.

\subsection{Energy and angular momenta}

Now we evaluate the conserved quantum numbers of the rotating membrane
solutions. 
We start 
with the action (\ref{actgen}),
\begin{align}
  S=&-\frac{T_2}{2}\int d^{3}\xi \Bigg(L(\dot{X_{0}}^{2} 
    -\dot{r_{5}}^2-\dot{r_{6}}^2 
    -\sum_{a=1}^4 (r_{a}^{2}\dot{\b_{a}}^{2}
    +\dot{r}_{a}^{2}\b_{a}^{2}))\nonumber\\ 
    &+\frac{1}{L}\{r_a,X_{0}\}^2 +\frac{1}{L}\sum_{a,b=1}^6
\Big(\frac{1}{2}\{r_{a},r_{b}\}^{2}
    +\frac{1}{2}(r_{a}r_{b})^{2}\{\b_{a},\b_{b}\}^{2}
    + r_{b}^{2}\{r_{a},\b_{b}\}^{2}\Big)\Bigg)\ ,
\end{align}
with coordinates defined as
\begin{align}
&Z_a(\tau,\sigma,\rho)=r_a(\tau,\sigma,\rho)e^{i\beta_a(\tau,\sigma,\rho)},
\quad a=1,...,4 \quad \nonumber\\
&X_9\equiv r_5\ ,\ \  \ X_{10}\equiv r_6\ ,\ \ \  X_0 \ .
\end{align}

The conserved quantum numbers are 
\begin{align}\label{EE}
E&=\int d\sigma d\rho ~ \Pi_{0}, 
\quad \Pi_{0}\equiv \frac{\delta S}{\delta \dot{X^{0}}}
\\\label{JJ}
J_{a}&= \int d\sigma d\rho ~ \Pi_{\beta_{a}},
\quad \Pi_{\beta_{a}}\equiv \frac{\delta S}{\delta \dot{\beta^a}}
\end{align}
In the present case, we obtain
\be
E=4\pi^2 T_{2}L~\kappa  \ ,\ \ \ \ \ 
J_a=4\pi^2 T_{2}L~\omega_{a}r_{a}^{2}\ .
\label{dre}
\ee
The energy of our solution can be
obtained from (\ref{cass}) (coming from 
the constraint (\ref{memcos2})), which determines $\kappa $,
\be
E^{2}=
2(4\pi^2 T_2 L)\sum_{a}J_{a}\w_{a}\ ,
\label{epo}
\ee
where $\w_{a} $ are determined by (\ref{consomega}).
Note that each term in the sum on the right hand side of (\ref{epo})
is positive definite, since $J_{a}\w_{a}= 4\pi^2 T_{2}L~\omega_{a}^2r_{a}^{2}$.
The constraints (\ref{consans1}) become
\be
\sum_{a=1}^4J_{a}k_{a}=0\ ,\ \ \ \ 
\sum_{a=1}^4J_{a}l_{a}=0\ .
\label{cffg}
\ee
Let us now  particular 
cases where some of the angular momenta vanish.
Because of the constraint  (\ref{consans1}), the 
minimum number of non-vanishing angular momenta is
two. However, in the case of rotation in two 
planes, i.e, $Z_3, Z_4 =const$, it can be seen that eqs. (\ref{consomega}), 
(\ref{consans1}) lead to $\w_1=\w_2=0$, so this solution is trivial. 
The first nontrivial case is the rotation in three planes.\footnote{
When the coordinates $X_9, X_{10} $ are compact, there are solutions
with rotation in one and two planes, described in section 2.4. In the case of
eleven uncompact coordinates, there is   a solution 
with rotation in two planes in the
 light-cone gauge, where the constraint (\ref{consans1}) is absent.
In section 5 we discuss the analogous solution in the Matrix theory.}

In the case of rotation in three planes, 
we can write (\ref{consomega}), 
(\ref{consans1}) as
\be
\sum_{a=1}^3J_{a}k_{a}=0\ ,\ \ \ \ 
\sum_{a=1}^3J_{a}l_{a}=0\ ,
\label{cff}
\ee
\be
\tilde{\w}_b^2-\sum_{a=1}^3\frac{J_a}{\tilde\w_a }(l_bk_a-l_ak_b)^2=0\ ,
\label{wer}
\ee
where we have defined $\tilde\w_a\equiv (4\pi^2 T_2 L^3)^{1/3}\w_a$.
In (\ref{wer}) we have expressed the radii in terms of the
angular momenta using (\ref{dre}). 
The reason is that 
the energy must be expressed in terms of the 
conserved quantum numbers, angular momenta
and winding numbers, characterizing the state
(winding numbers are not conserved in interactions because
they are along contractible circles; nevertheless, they 
characterize the states of the free theory). 
The parameters  $\tilde\w_a $ are
dimensionless (just as $\w_a $), but they depend only on $J_a, l_a, k_a$
through (\ref{wer}). 
Equations (\ref{wer}) are  coupled cubic equations, and the analytic
solution is complicated. 
It is easy to see that  solutions
with real frequencies (and real radii) exist. 
For example, with the choice
\be
(k_1,k_2,k_3)=(1,-3,-4)\ ,\ \ \ \ 
(l_1,l_2,l_3)=(4,3,-11)\ ,\ \ \ \ 
(J_1,J_2,J_3)=(4,2,2)\ ,
\ee
the constraints (\ref{cff}) are satisfied, and from (\ref{wer}) we
obtain the real solution   
$(\tilde{\w}_1,\tilde{\w}_2,\tilde{\w}_3)
=(4.9,21.5,14.4)$
(for which also the radii are real, since $J_a/\w_a $ are positive, 
see (\ref{dre})).
In conclusion, the energy in the  case of rotation in three planes is 
\be
E^2=2(4\pi^2 T_2)^{2/3}\sum_{a=1}^{3}J_{a}\tilde \w_{a}\ ,
\ee
where the $\tilde \w_{a}$ are determined  by eq.~(\ref{wer}) in 
terms of integer quantum numbers 
$k_a,l_a,J_a$  obeying (\ref{cff}).
As expected,  the energy does not depend on the arbitrary length
parameter $L$ introduced
in the choice of gauge.

\subsection{More general rotating solutions}

We now consider the membrane 
wrapped around two compact directions, 
$X^9$ and $X^{10}$, with linear momentum,
\bea
X_9 &=& R_9(n_9\sigma+m_9\rho)+q_9 \tau\ ,
\nonumber \\
X_{10} &=& 
R_{10}(n_{10}\sigma+m_{10}\rho)+q_{10}\tau \ .
\label{xxop}
\eea
$$
$$
For the coordinates $Z_a$, $a=1,...,4$, we use the same ansatz
(\ref{Za}).
 The constraints (\ref{memcos1}), (\ref{memcos2}) now take the form
\begin{align}
&\sum_{a=1}^4\omega_ar_a^2k_a+R_9q_{9}n_{9}
+R_{10}q_{10}n_{10}=0\ ,
\\
&\sum_{a=1}^4\omega_ar_a^2l_a+R_9q_{9}m_{9}
+R_{10}q_{10}m_{10}=0\ ,
\\\label{genK}
&L^{2}(\kappa^2-\sum_a\omega_a^2r_a^2-q_9^2-q_{10}^2)=
L^{2}\sum_{a=1}^{d} r_{a}^{2}\w_{a}^{2} +R_{9}^{2}R_{10}^{2}(n_{9}m_{10}
-n_{10}m_{9})^{2}\ ,
\end{align}
where in the last equation we have used the 
relation
\be\label{tilw} 
\tilde{\w}_c^2-\sum_a\frac{J_a^2}{\tilde{\w_{a}}}(l_ck_a-l_ak_c)^2
-(4\pi^2T_2)^{2/3}R_{9}^{2}(m_{9}k_{c}-n_{9}l_{c})^{2}
-(4\pi^2T_2)^{2/3}R_{10}^{2}(m_{10}k_{c}-n_{10}l_{c})^{2}=0,
\ee
which follows from  the equation of motion for the radial
coordinates. The parameters $\tilde{\w}_a$ are defined as before,
 $\tilde{\w}_a \equiv (4\pi^2T_2L^3)^{1/3}\w_a$.

The linear momenta along $X_{9}$ and $X_{10}$ are 
\be
P_{i}=\int d\sigma d\rho ~ \Pi_{i},
\quad \Pi_{i}\equiv \frac{\delta S}{\delta \dot{X^{i}}}\ .
\ee
We obtain
\be
P_9=4\pi^2 T_2 L q_{9}\ ,\ \ \ \ P_{10}=4\pi^2 T_2 L q_{10} \ .
\ee
Because $X_9$ and $X_{10}$ are compact coordinates, they are
quantized, 
\be
P_{9}\equiv \frac{\tilde{n}_{9}}{R_{9}}\ ,
\quad P_{10}\equiv \frac{\tilde{n}_{10}}{R_{10}}\ .
\ee
The constraints take the form
\be
\sum_{a=1}^4J_{a}k_{a}+ n_9\tilde{n}_9+n_{10}\tilde{n}_{10}=0\ ,\ \ \ \ 
\sum_{a=1}^4J_{a}l_{a}+ m_9 \tilde{n}_9+m_{10}\tilde{n}_{10}=0\ .
\label{cffy}
\ee
The energy is then
obtained from (\ref{dre}) with $\kappa $ determined by
 (\ref{genK}). We get
\be\label{Eradi}
E^{2}=
2(4\pi^2T_{2})^{2/3}\sum_{a=1}^4 J_{a}\tilde{\w}_{a}
+(4\pi^2T_{2})^2R_{9}^{2}R_{10}^{2}(n_{9}m_{10}-n_{10}m_{9})^{2}
+\frac{\tilde{n}_{9}^{2}}{R_{9}^2}+\frac{\tilde{n}_{10}^{2}}{R_{10}^2}\ .
\ee
The $\tilde{\w}_{a}$ are numbers  determined  by (\ref{tilw}) in terms of the 
integer
 quantum numbers $J_a, l_c$, $k_c$, $m_9$, $n_9$, $m_{10}$, $n_{10}$ and the 
parameters $R_9, R_{10}$. Note that in the energy $E$
there is no dependence on the arbitrary constant
$L$ as expected.
The second term proportional to $R_9^2R_{10}^2$ is the usual contribution
to the energy coming from the torus area, times membrane tension, times
membrane charge (equal to $n_{9}m_{10}-n_{10}m_{9}$).

\subsubsection{Rotation in one plane}

The simplest rotating solution is the case  $Z_2,Z_3,Z_4=$const.
corresponding to rotation in one plane $Z_1$, 
with the previous ansatz for $X_9,X_{10}$,
\begin{align}
X_0&=\kappa \tau   \ ,\nonumber\\
Z_{1}&=re^{i\w\tau + i(k \sigma + l \rho)} \ ,\nonumber\\
X_9 &=R_9(n_9\sigma+m_9\rho)+q_9 \tau \ ,
\nonumber \\
X_{10} &=
R_{10}(n_{10}\sigma+m_{10}\rho)+q_{10}\tau \ .\nonumber
\end{align}
In this case, the constraints become, 
\begin{align}
  &kJ + n_9\tilde{n}_9+n_{10}\tilde{n}_{10}=0\ ,
\nonumber\\
  &lJ + m_9 \tilde{n}_9+m_{10}\tilde{n}_{10}=0\ ,
\nonumber\\
  &L^2\w^2=R_{9}^{2}(m_{9}k-n_{9}l)^{2}
  +R_{10}^{2}(m_{10}k-n_{10}l)^{2}\ .
\nonumber
\end{align}
The energy is then
\begin{align}
E^2&=2(4\pi^2T_{2}L) J\w
+(4\pi^2T_{2})^2R_{9}^{2}R_{10}^{2}(n_{9}m_{10}-n_{10}m_{9})^{2}
+\frac{\tilde{n}_{9}^{2}}{R_{9}^2}+\frac{\tilde{n}_{10}^{2}}{R_{10}^2}
\nonumber\\
&=\left((4\pi^2 T_2)(n_9m_{10}-n_{10}m_{9})R_9R_{10}
+ \frac{1}{R_{9}}\sqrt{\tilde{n}_9^2
+\tilde{n}_{10}^2\frac{R_9^2}{R_{10}^2}}\ \right)^2\ .
\label{onep}
\end{align}
Thus the energy is a complete square. This is related to the
fact that the corresponding quantum state is a BPS state.
In fact, it is the same BPS state studied in \cite{schwarz,waves}.
In general, the BPS state of  \cite{schwarz,waves} 
represents a non-marginal bound state
of fundamental string and D string in type IIB string theory, 
with charges $\tilde n_9, \ \tilde n_{10}$
and momentum $n\equiv n_9 m_{10}-n_{10}m_9$ (see also section 3.2).
It describes an excited string state (or excited membrane state), 
whose specific quantum
state can be any of the exponential number of states at this level.
The present solution represents one of these quantum states, namely the state
with maximum angular momentum.

\subsubsection{Rotation in two planes}

Let us now consider the solution describing rotation in two planes
$Z_1, Z_2$, by setting $Z_3=Z_4=$const.
The constraints and the equations of motion become, 
\begin{align}
  &k_1J_1 + k_2J_2+ n_9\tilde{n}_9+n_{10}\tilde{n}_{10}=0\ ,
\nonumber\\
  &l_1J_1 +l_2J_2+ m_9 \tilde{n}_9+m_{10}\tilde{n}_{10}=0\ ,\nonumber\\
  &\tilde{\w}_1^2=\frac{J_{2}}{\tilde{\w}_2}(k_1l_2-k_2l_1)^2
  +(4\pi^2T_2)^{2/3}\bigg[R_{9}^{2}(m_{9}k_1-n_{9}l_1)^{2}
  +R_{10}^{2}(m_{10}k_1-n_{10}l_1)^{2}\bigg]\ ,
\nonumber\\
  &\tilde{\w}_2^2=\frac{J_{1}}{\tilde{\w}_1}(k_1l_2-k_2l_1)^2
  +(4\pi^2T_2)^{2/3}\bigg[R_{9}^{2}(m_{9}k_2-n_{9}l_2)^{2}
  +R_{10}^{2}(m_{10}k_2-n_{10}l_2)^{2}\bigg]\ .
\nonumber
\end{align}
The solution simplifies when the linear momenta along the directions
 coordinates $X_9$, $X_{10}$ vanish, i.e. $\tilde n_9=\tilde
 n_{10}=0$. In this case, we have
\be
J_1k_1 + J_2k_2=J_1l_1+J_2l_2=0\ ,
\label{JL}
\ee
so that $l_1k_2=l_2k_1$, and 
we find an explicit simple expression for the frequencies
\begin{align}
\label{ww}
&\tilde{\w}_1^2=(4\pi^2T_2)^{2/3}\bigg[R_{9}^{2}(m_{9}k_1-n_{9}l_1)^{2}
  +R_{10}^{2}(m_{10}k_1-n_{10}l_1)^{2}\bigg]\\
\label{wwq}
  &\tilde{\w}_2^2=(4\pi^2T_2)^{2/3}\bigg[R_{9}^{2}(m_{9}k_2-n_{9}l_2)^{2}
  +R_{10}^{2}(m_{10}k_2-n_{10}l_2)^{2}\bigg]\ .
\end{align}
Then, the energy of this solution is
\begin{align}
E^2=&2(4\pi^2T_{2})^{2/3}\sum_{a=1}^{2}J_a\tilde{\w}_a
+(4\pi^2T_{2})^2R_{9}^{2}R_{10}^{2}(n_{9}m_{10}-n_{10}m_{9})^{2}
\nonumber\\
=&2(4\pi^2T_{2})\sqrt{R_9^2\lp\frac{k_1}{l_1}m_9-n_9\rp^2+R_{10}^2\lp
\frac{k_1}{l_1}m_{10}-n_{10}\rp^2}\ (|l_1J_1|+|l_2J_2|)
\nonumber\\
&+(4\pi^2T_{2})^2R_{9}^{2}R_{10}^{2}(n_{9}m_{10}-n_{10}m_{9})^{2}\end{align}
We have added absolute value bars to the two terms in the sum 
to account for the fact that
each term in the sum $\sum_a J_a\tilde\w_a $ is positive definite because  
$J_a\sim  r^2_a\tilde\w_a$.
Note that, unlike the previous case, 
the energy is not a complete square. 

In the case $X_{9}=n_9R_9\sigma$ and
 $X_{10}=R_{10}\rho$, i.e. $m_9=n_{10}=0,\ m_{10}=1$, the energy becomes
\begin{align}\label{emain}
E^2&=2(4\pi^2T_{2})\sqrt{R_9^2n_9^2+R_{10}^2\frac{k_1^2}{l_1^2}}\left(|l_1J_1|
+|l_2J_2|\right)
+(4\pi^2T_{2})^2R_{9}^{2}R_{10}^{2}n_{9}^2\ .\nonumber\\
\end{align}
We will return to this energy formula in the next section.


   \setcounter{equation}{0}
\section{Non-perturbative states in type II string theory  from
  rotating membranes}

\subsection{Type IIA string theory or M-theory on $S^1$}

Type IIA string theory is obtained from M-theory by compactifying
the eleventh dimension $X^{10}$ on a circle.
The string tension and string coupling are related to the
membrane and M-theory parameters as follows:
\be
\alpha'=\frac{1}{4\pi^2R_{10}T_2}, \qquad g^2_{\rm IIA}=4\pi
^2R_{10}^3 T_2\ ,
\ee
\be
T_1=(2\pi\alpha ')^{-1}=2\pi R_{10}T_2 \ ,
\ee
\be
\kappa_{11}^2=16\pi^5 l_{P}^9\ ,\ \ \ \ T_2=(2\pi
l_{P}^3)^{-1}=(4\pi^2{\alpha ' }^{3/2} g_{\rm IIA})^{-1}\ ,\ \ \  \ 
R_{10}^2= \alpha'  g^2_{\rm IIA}\ .
\ee
The perturbative solutions of type IIA string theory have the  form dictated by
the ``double
dimensional reduction'' ansatz \cite{duffd}, 
\be
X^{10}=R_{10}\rho \qquad \partial_{\rho}X^{\mu}=0, 
\quad \mu=0,\ldots,9\ .
\label{jua}
\ee
The membrane solutions of the previous sections have explicit
dependence
on the $\rho $ coordinate, and also momentum in the eleventh coordinate
$X^{10}$. One expects that in string theory they arise as
non-perturbative objects.

First, consider the type IIA limit in ten uncompact dimensions 
which is achieved by setting $R_9\to \infty $. In this case we must
set $n_9=m_9=0$ in the solution (\ref{Za}), (\ref{xxop}). Now 
the momentum $P_9$  is continuous.
In addition, we set $n_{10}=0,\ m_{10}=1$ 
to have $X^{10}=R_{10}\rho +q_{10}\tau  $. 
In terms of the type IIA parameters, the energy (\ref{Eradi}) of the membrane
solution of section 2.4 then takes the form
\begin{align}
E^2=P_9^2+\frac{2}{\alpha'g_{\rm IIA}^{2/3}}\sum_{a=1}^4J_a
\ \tilde{\w}_a(g_{\rm
  IIA})+ 
\frac{\tilde{n}_{10}^2}{\alpha'g_{\rm IIA}^2}\ ,
\label{kko}
\end{align}
where angular momenta and winding numbers satisfy the constraints
\be
\sum_{a=1}^4 k_a J_a =0\ ,\ \ \ \ \sum_{a=1}^4 l_a J_a =-\tilde n_{10}\ .
\label{conh}
\ee
The $\tilde{\w}_a$ are determined in terms of the coupling constant $g_{\rm
  IIA}$ and the conserved quantum numbers 
by the system of equations (\ref{tilw}), 
which in terms of string parameters reads
\be
\tilde{\w}_c^2-\sum_a\frac{J_a^2}{\tilde{\w_{a}}}(l_ck_a-l_ak_c)^2
-g_{\rm IIA}^{4/3} k_{c}^{2}=0\ .
\ee
In order to solve explicitly the above equations, we consider again
the case of 
rotation in two planes. Then the equations for the frequencies are given by
\bea
\tilde \omega_1^2-\frac{J_2^2}{\tilde \omega_2}\lp
l_1k_2-k_1l_2\rp^2
=g_{\rm IIA}^{4/3}k_1^2 \ , \nonumber \\
\label{eqstwoplanes}
\tilde \omega_2^2-\frac{J_1^2}{\tilde \omega_1}\lp
l_1k_2-k_1l_2\rp^2
=g_{\rm IIA}^{4/3} k_2^2 \ .
\eea
These equations admit a  number of solutions,
which simplify in the particular case of spins of equal magnitude
and opposite sign,
$J_1=-J_2$ (which, by the constraint equations (\ref{conh}), 
implies $k_1=k_2$). 
In this case the frequencies 
are given by
\be
\tilde{\omega}_1=\frac{2g_{\rm IIA}^{8/3}k_1^2}{\tilde{n}_{10}^2}\frac{1}{1
+\sqrt{1+\frac{4g_{\rm IIA}^4k_1^2}{\tilde{n}_{10}^4}}} \ ,
\ee
\be
\tilde{\omega}_2=-\frac{\tilde{n}_{10}^2}{2g_{\rm IIA}^{4/3}}
\lp 1
+\sqrt{1+\frac{4g_{\rm IIA}^4k_1^2}{\tilde{n}_{10}^4}} \rp\ .
\ee
Inserting into (\ref{kko}), we obtain the exact expression for the
energy of this configuration.
At weak coupling,
$g_{\rm IIA}\ll 1$, the  energy has the following expansion 
\be
\alpha'E^2=\frac{2J_1\tilde{n}_{10}^2}{g_{\rm IIA}^2}+4J_1g_{\rm IIA}^2
\frac{k_1^2}{\tilde{n}_{10}^2}+\mo\lp g_{\rm IIA}^{6}\rp \ .
\ee
There are several points which are worth noticing.
First, we again find a Regge-type formula. A priori, this is not  obvious
from (\ref{kko}), since the frequencies in general depend on the
angular momenta. The point is that  the frequencies depend
on the combination $J_{1,2}^2(l_1k_2-l_2k_1)^2$ which for $J_1=-J_2$
is equal to $\tilde n_{10}^2$ (see  (\ref{conh})).
Second, also for this solution the energy has the behavior 
$E={\rm const.} {1\over g_{\rm IIA}}$, characteristic of D branes. As explained
in the introduction, this
guarantees that one can ignore gravitational back reaction effects in
the weak coupling limit.
The origin of the  $1/g_{\rm IIA}$ behavior is a D0 brane charge,
coming from  momentum of the membrane in the $X_{10}$ direction.
Because the membrane has zero charge, now there is no D2 brane and no
winding charge for the fundamental string.
Thus the state  represents a rotating system of D0 brane and
fundamental string in uncompactified ten-dimensional type IIA string
theory.

\medskip

Now we consider the case of compact $X_9$ coordinate, and the solution 
with energy given by (\ref{emain}).
In terms of type IIA string theory parameters, the analytical energy formula 
(\ref{emain}) --describing rotation in two planes--
becomes
\be
E^2=\frac{2}{\a'}\sqrt{\frac{R_9^2n_9^2}{\a'g_{\rm IIA}^2}+\frac{k_1^2}{l_1^2}}
\ (|l_1J_1| + |l_2J_2|)+\frac{ R_9^2n_9^2}{\a'^2}\ ,
\label{masso}
\ee
$$
k_1J_1+k_2J_2=l_1J_1+l_2J_2=0\ .
$$
At weak coupling, the energy behaves as $E={\rm const.} {1\over g_{\rm IIA}}$.
This behavior is characteristic of D branes. For this particular 
solution
there is no D0 brane present, since we have set the D0 brane charge
$\tilde n_{10}$ to zero. The origin of the  $1/g_{\rm IIA}$ behavior is
the presence of a D2 brane. Indeed, for this solution, 
the membrane charge is equal to
$n_9m _{10}-n_{10}m_9=n_9$. Since the membrane is also extended in the
$Z_1, \ Z_2$ and $X_{10}$ directions, 
after dimensional reduction one is left with a rotating bound state of
a D2 brane  and a fundamental string (with winding charge also
equal to $n_9$). Remarkably, the energy formula has a simple
Regge-type 
behavior, $E^2\sim J$.

Now consider the particular case $n_{9}=0$. We get
\begin{align}
E^2=2(4\pi^2T_{2}R_{10})\left(|k_1J_1|
+|k_2J_2|\right)=\frac{2}{\alpha'}\left(|k_1J_1| + |k_2J_2|\right)
\label{pertu}
\end{align} 
It is interesting that the energy of these states does not depend on
$l_1, l_2$. In particular, the energy is the same as in the case
$l_1=l_2=0$, which in the dimensionally reduced theory  corresponds 
to a  string (see (\ref{Za}), (\ref{jua})). 
Thus there is a  family of
membranes (parametrized by $l_1$, with $l_2=-l_1J_1/J_2$) 
giving rise to states  with the same energy in the string theory.

The  solutions with $l_1=l_2=0$ and energy given by (\ref{pertu})
are the solutions of \cite{CIR}.
These solutions were found to be quantum mechanically very stable, with
a lifetime  proportional to \cite{CIR} $g^{-2}_{\rm IIA}$(mass)$^5$.
The reason of this long life time is that the closed string cannot
classically break
 due to the fact that during the evolution 
there is never contact
between two points of the string. It can only decay by
emitting light modes. These states are probably the most stable states in
the string spectrum. 
We expect that the present
membrane solutions with $n_9\neq 0$ and energy given by 
(\ref{masso}) are also long-lived quantum mechanically.

An interesting question is what is the supergravity solution
describing
these spinning membranes. Spinning M-brane supergravity solutions were
given in \cite{cvetic}. However, none of these solutions 
describe the present configurations.
The reason is that the present membrane configurations  
rotate along the directions where the membrane is extended.

\subsection{Type IIB string theory}

Type IIB string theory arises by compactifying M theory on the torus
$X_9, X_{10}$. The  type IIB parameters are related to M theory
parameters as follows:
\be
\alpha'=\frac{1}{4\pi^2R_{10}T_2}\ , \qquad 
g^2_{\rm IIB}=\frac{R_{10}^2}{R_{9}^{2}}\ , \qquad R_{9B}=\frac{\a'}{R_9}\ .
\ee
In terms of type IIB parameters, 
the general expression of the energy  (\ref{Eradi}) reads
\begin{align}
E^{2}=
2\lp\frac{R_{9B}}{g_{\rm IIB}\alpha'^2}\rp^{2/3}\sum_{a=1}^4
 J_{a}\tilde{\w}_{a}
+\frac{1}{R_{9B}^2}(n_9m_{10}-n_{10}m_9)^2
+\frac{R_{9B}^2}{\a'^2}\left(\tilde{n}_9^2+\frac{\tilde{n}_{10}^{2}}{g_{\rm
    IIB}^2}\right)\ .
\end{align} 

Let us now consider particular cases. In the case 
of rotation on one plane, eq.~(\ref{onep}), we have
\be
E^2=\left( (n_9m_{10}-n_{10}m_{9}){1\over R_{9B}}
+ \frac{ R_{9B}  }{\alpha' }\sqrt{\tilde{n}_9^2
+\frac{\tilde{n}_{10}^2}{g_{\rm IIB}^2}}\ \right)^2\ ,
\ee
describing the BPS non-marginal bound state of D string 
(charge $\tilde n_{10}$),
fundamental string (charge $\tilde n_9$) and momentum 
$n\equiv n_9m_{10}-n_{10}m_{9}$, as anticipated in section 2.4.1.

In the case of the solution   (\ref{emain}) describing rotation in 
two planes, we have
\be
E^2=\frac{2}{\a'}\sqrt{\frac{n_9^2}{g_{\rm IIB}^2}+
\frac{k_1^2}{l_1^2}}\ (|l_1J_1|
+|l_2J_2|) + \frac{n_9^2}{R_{9B}^2} \ .
\ee
This is the T-dual of the D2 brane/fundamental string system described
in section 3.1. Because the D2 brane is extended in $X_9$ and $Z_1,
Z_2$ directions, T-duality in $X_9$ should produce a complicated
rotating system 
involving 
D-strings, D3 branes  and fundamental strings, with momentum $n_9$.
It is remarkable that the energy of such a system is given by a simple
Regge-type  formula. As the type IIA T-dual counterpart, we expect that
this system is classically stable and long-lived quantum mechanically.

  \setcounter{equation}{0}
\section{Spinning membranes in $AdS_p\times S^q$}

In this section we  construct analogous membrane solutions
in a curved background spacetime, $AdS_p\times S^q$, which
are relevant for AdS/CFT correspondence applications \cite{malda}.
First we obtain the general expression for our solutions, which are
of the form  (\ref{Za}), and then we find an explicit
expression for the energy of these solutions in the large
angular momentum limit. The solutions generalize the rotating circular
strings of \cite{ART} to membrane theory.

\medskip

We consider membranes on  $AdS_7\times S^4$ and $AdS_4\times
S^7$ or, generically, $AdS_p\times S^q$.
 Let
$Y^{\mu},\mu=0,\ldots,p$ be the embedding coordinates in the AdS space and
$X^k,k=1,\ldots,q+1$ the embedding coordinates in the sphere. 
 The action for the membrane reads
\bea
S=\frac{T_2}{2}\int d^3\xi\Bigg(
-\sqrt{-h}h^{\alpha\beta}\lp \partial_{\alpha}
Y^{\mu}\partial_{\beta}
Y^{\nu}\eta_{\mu\nu}+ \partial_{\alpha}
X^{k}\partial_{\beta}
X^{k}\rp+\frac{1}{2}\sqrt{-h} \nonumber \\
\label{adsaction}
+\ \tilde{\Lambda}\lp 
Y^{\mu}Y^{\nu}\eta_{\mu\nu}+R_A^2\rp+
\Lambda\lp 
X^{k}X^{k}-R_S^2\rp\Bigg) \ .
\eea
The Lagrange multipliers enforce the constraints
\be
\sum_{k=1}^{q+1}X^2_k=R_S^2, \qquad Y^{\mu}Y^{\nu}\eta_{\mu\nu}=-R_A^2
\ ,
\ee
where $\eta_{\mu\nu}=(-,+,\ldots,+,-)$. We have defined
$\tilde{\Lambda}, \ \Lambda$ so that they transform as $\sqrt{-h}$
  under world-volume reparametrizations.
The radii of the sphere and of the AdS space are 
given by
\bea R_{S}=2R_{AdS}=l_{P}(2^5\pi^2N)^{1/2}, \quad (AdS_4\times S^7)\ ,
\nonumber
\\
\label{radiival}
R_{S}=R_{AdS}/2=l_{P}(\pi N)^{1/3}, \quad (AdS_7\times S^4)  \ ,
\eea
where $l_{P}$ is the Planck length.
From the action (\ref{adsaction}) one  derives the following equations of
motion
for the coordinates, 
\be
\Lambda X_k=\partial_{\beta}\lp
\sqrt{-h}h^{\alpha\beta}\partial_{\alpha}X_k
\rp \ ,
\ee
\be
\tilde{\Lambda} Y_{\mu}=
\partial_{\beta}
\lp \sqrt{-h}h^{\alpha\beta}\partial_{\alpha}Y_{\mu}\rp \ .
\ee
Now we consider the following ansatz,
\bea
Z_1\equiv X_1+iX_2=r_1e^{ i\omega_1\tau +im_1\sigma+in_1\rho }
\nonumber \ ,\\
\label{adsan}
Z_2\equiv X_3+iX_4=r_2e^{ i\omega_2\tau +im_2\sigma+in_2\rho}  \ ,
\\
\vdots \nonumber \\
Z_d\equiv X_{2d-1}+iX_{2d}=r_de^{i\omega_d\tau +im_d\sigma+in_d\rho}
\nonumber  \ ,
\eea
where $d=2$ for $S^4$ and $d=4$ for $S^7$.
For the AdS coordinates, we take
\be
Z_0\equiv Y_0+iY_p=R_A e^{ i\omega_0\tau} \ .
\ee
We use the 
same gauge as in  section 2, given by $
h_{\tau\sigma}=0$, $h_{\tau\rho}=0$ and $ h_{\tau\tau}L^2=-g$. 
For our ansatz   (\ref{adsan}), these constraints read 
\be
\label{con1}
\sum_{a=1}^dm_a\omega_ar_a^2= \sum_{a=1}^dn_a\omega_ar_a^2=0 \ ,
\ee
\be
\label{con2}
R_A^2\omega_0^2-\sum_{a=1}^dr_a^2\omega_a^2=\frac{1}{L^2}
\sum_{b<a}r_a^2r_b^2(m_bn_a-m_an_b)^2 \ .
\ee
The equations of motion 
that are derived from the action  (\ref{adsaction}) 
give rise to the following relations between the parameters: 
\be
\label{eom21}
\tilde{\Lambda}=-L\omega_0^2 \ ,
\ee
\begin{align}
-\Lambda&=L\omega_a^2-\frac{1}{L} h_{\rho\rho}m_a^2-
\frac{1}{L} h_{\sigma\sigma}n_a^2+
\frac{2}{L} h_{\sigma\rho}m_an_a \nonumber\\
&=Lw_a^2 - \frac{1}{L}\sum_{b=1}^d r_b^2\left(n_bm_a-n_am_b\right)^2\ .
\label{eom22}
\end{align}
The energy and the angular momenta can be
derived from (\ref{EE}), (\ref{JJ}),
\def\J{ \mathcal{J}}
\def\E{ \mathcal{E}}
\be
J_a=4\pi^2T_2 L r_a^2\omega_a\equiv 4\pi^2 T_2L ~R_S^2\J_a \ ,
\ee
\be
\label{adsenergy}
E=4\pi^2 T_2LR_A\omega_0\equiv 4\pi^2 T_2LR_A~\E \ .
\ee
The
constraint $\sum_{a=1}^d r_a^2=R_S^2$ can be written as
\be
\label{radiicons}
\sum_{a=1}^d\frac{\J_a}{w_a}=R_S^2 \ .
\ee
Solving  (\ref{eom22}), one finds   the frequencies in terms of
$\Lambda $. Then $\Lambda $ is determined using
(\ref{radiicons}).
Plugging the result into (\ref{con2}) (with $r_a^2=R_S^2\J_a/\omega_a
$),
one determines the energy 
(\ref{adsenergy}) in terms of $J_a$ and winding numbers.
The system of equations can be  solved systematically 
as power series in  $\frac{1}{\J}$,
\begin{align}
\J\equiv\sum_{a=1}^d \J_a \ .
\end{align}
Note that large $\J$  implies also large Lagrange multiplier $|\Lambda|$.
Indeed, since
the radii $r_a$ on the sphere are bounded, 
large angular momenta requires
large frequencies, which, by (\ref{eom22}), implies large
$|\Lambda|$, and the scaling $|\Lambda|\sim \J^2$.
This situation is similar to that of
spinning strings, see \cite{ART}. 
So we first solve equation (\ref{eom22}) for the frequencies 
(inserting $r_a^2=R_S^2\J_a/\omega_a
$) using 
perturbation theory and then plug the expression of the frequencies in the
constraint (\ref{radiicons}) to express $\Lambda$ 
in terms of the angular momenta
and the winding numbers. 
We find
\begin{align}
w_c&=\J+\frac{R_S^2}{2L^2\J^2}
\sum_{a=1}^{d}\J_a(n_cm_a-n_am_c)^2+\mathcal{O}\lp\frac{1}{\J^2}\rp\ , 
\\
|\Lambda|^{1/2}&=\J-\frac{R_S^2}{2L^2\J^3}
\sum_{a,c}^{d}\J_a\J_c(n_cm_a-n_am_c)^2 +\mathcal{O}\lp\frac{1}{\J^2}\rp
 \ .
\end{align}
Hence we find the energy:
\begin{align}\label{rpp}
\E^2= \omega_0^2=\frac{R_S^2}{R_A^2}\J^2+\frac{R_S^4}{\J^2R_A^2L^2}
\sum_{a,b=1}^d\lp
m_an_b-n_am_b\rp^2\J_a\J_b+\mathcal{O}\lp\frac{1}{\J^2}\rp \ .
\end{align}
Taking the square root and expanding, we get
\be
\label{nloenergy}
\E-\frac{R_S}{R_A}\J=
\frac{R_S^3}{2\J^3R_AL^2}\sum_{a,b=1}^d (m_an_b-n_am_b)^2\J_a\J_b
+\mathcal{O}\lp\frac{1}{\J^2}\rp \ .
\ee
Finally, expressing eq. (\ref{nloenergy}) in terms of the physical 
energy and spin, one obtains that the dependence on the arbitrary constant $L$
cancels, as expected, to give the  result,
\be
E-\frac{J}{R_S}=
\frac{R_S^5(4\pi^2T_2)^2}{2J^3}
\sum_{a,b=1}^d (m_an_b-n_am_b)^2J_aJ_b
+\mathcal{O}\lp\frac{1}{J^2}\rp \ ,
\label{euo}
\ee
supplemented with the constraints (\ref{con1})
\be
\sum_{a=1}^d J_a m_a =\sum_{a=1}^d J_a n_a =0\ .
\ee
This spinning toroidal membrane solution is the  
analog of the circular spinning strings 
in $AdS_5\times S^5$ 
found in \cite{ART}. For the spinning strings, in the large angular
momentum limit, one finds the formula
\be
\E-\J=\frac{1}{2\J^2}\sum_{i=1}^3m_i^2\J_i+... \ .
\label{stra}
\ee
As in the case of $AdS_5\times S^5$ (\ref{stra}), 
the energy (\ref{euo}) of configurations with angular momentum
on the sphere exhibits the behavior $E\sim J$, as opposed to the Regge
behavior $E^2\sim J$ of flat space.
In the framework of the AdS/CFT correspondence, the energy formula
(\ref{euo}) gives the anomalous dimension
of a CFT operator with $SO(5)$ (for the 5+1 CFT) or $SO(8)$ 
for the 2+1 CFT) charges $J_a$.

\medskip

In the particular case of rotation in two planes,
there is 
a significant simplification, because in this case the
constraint (\ref{con1}) implies that $n_1m_2-m_2n_1=0$. 
As a result, the energy is given by the leading order term
without higher order corrections,
\be
E=\frac{J}{R_S} \ ,
\ee
This suggests that this configuration 
should be supersymmetric and that 
the corresponding CFT operator should be BPS,
since its bare dimension is exact in the classical membrane approximation.

  \setcounter{equation}{0}

\section{Membranes and Matrix theory}

In this section we consider solutions of the matrix
model equations which are the analogues of the classical membrane solutions
of Section 2.2, and study their properties.

\subsection{Membranes solutions and matrix model solutions}

The starting point is the bosonic part of the supersymmetric 
quantum-mechanics Yang-Mills Lagrangian describing the dynamics of $N$
D0 branes, given by
\be
L=\frac{1}{2R_{10}}\lp \mathrm{tr}\lc \dot{X}^i \dot{X}^i+
\frac{R_{10}^2}{2l_P^6}
\lc X^i,X^j\rc^2\rc\rp, \quad i=1,\ldots,8 \ ,
\label{mlag}
\ee
where $X^i$ are matrices of the Lie algebra of $SU(N)$. 
The equations of motion are
\be
\frac{d^2}{dt^2}X^k=- \frac{R_{10}^2}{l_P^6} 
\sum_{i=1}^8 \lc\lc X^k,X^i\rc,X^i\rc \ .
\ee
 This equation of
motion
is supplemented with the self-consistency Gauss constraint,
\be
\sum_{i=1}^8 \lc X_i,\dot{X}_i\rc=0 \ .
\ee
In order to establish a dictionary between membrane
solutions and matrix model solutions, let us recall 
how the toroidal membrane arises as the $N=\infty$ limit of the
$SU(N)$
matrix model. The toroidal membrane coordinates
can be expanded as
\be
\label{general}
X^{i}\lp\tau,\sigma,\rho\rp=\sum_{n,m=-\infty}^{\infty}
X_{nm}^{i}(\tau)e^{in\sigma+im\rho}\equiv \sum_{n,m=-\infty}^{\infty}
X_{nm}^{i}(\tau)T_{nm}\ ,
\ee
where $T_{n_1n_2}\equiv T_{\vec{n}}$ are  generators of the
area-preserving diffeomorphism algebra of the torus,
\be
\{ T_{\vec{n}},T_{\vec{m}} \} =(\vec{n}\times
\vec{m})T_{\vec{n}+\vec{m}}\ ,\ \ \ \ \ 
(\vec{n}\times \vec{m})\equiv
n_1m_2-m_1m_2\ .
\label{toro}
\ee
The analogue of the $T_{\vec{n}}$ generators in the $SU(N)$ matrix
 model
 is a special basis of generators $\{J_{\vec{n}}\}$ of  the Lie
 algebra of $SU(N)$  satisfying the algebra
\cite{infinite,floratos2}
\be
\lc J_{\vec m},J_{\vec n}\rc=-2i\sin \lp {2\pi\over N}
(\vec{n}\times \vec{m})\rp
J_{\vec{m}+\vec{n}}\ ,\ \ \ \ N={\rm odd}\ ,
\label{alfaro}
\ee
\be
\lc J_{\vec m},J_{\vec n}\rc=-2i\sin \lp {\pi\over N}
(\vec{n}\times \vec{m})\rp
J_{\vec{m}+\vec{n}}\ ,\ \ \ \ N={\rm even}\ .
\label{alfar}
\ee
 In the Appendix we review the construction and properties
of these algebras.
A general matrix on the Lie algebra can be expanded as
\be
X^{i}\lp\tau \rp=\sum_{n,m=0}^{N-1}{}'~
X_{nm}^{i}(\tau)~ J_{nm}\ ,
\label{jui}
\ee
where prime means excluding the term $(n,m)=(0,0)$.
In the  $N=\infty$ limit, the algebra (\ref{alfaro}),  (\ref{alfar})
approaches the area-preserving
diffeomorphism algebra of the torus, and the $SU(N)$ matrix theory
is expected to reproduce exactly the same dynamics of membrane theory
in the light-cone gauge, since they are described by the same Hamiltonian.

Given any toroidal membrane classical solution, it can be expanded as
in (\ref{general}). Then one can write down an ansatz
for a matrix model solution of the form (\ref{jui})
 by taking the same coefficients $X_{nm}^{i}$ and extending
the sum taking into account the periodicity,
$J_{(m_1+Nk_1,m_2+Nl_1)}=J_{(m_1,m_2)}$. For the rotating membrane
solutions of section 2, the ansatz is straightforward since in complex
coordinates $Z_a$ there is a single term  in the sum
(\ref{general}), i.e. $Z_a= r_a e^{i\omega_a \tau}\  e^{ik_a\sigma+il_a \rho}$.

\subsection{Rotating matrix model solution}

Here we review the  rotating solution
of the matrix model given in \cite{AFP}. The starting point
is the matrix model
Lagrangian (\ref{mlag}) in complexified coordinates as
\be
L=\frac{1}{2R_{10}}{\rm tr}\big[ \dot{Z_a}\dot Z_a^{\dag}\big]+
\frac{R_{10}}{8l_P^6}
{\rm tr} \bigg(\lc Z_a,Z_b\rc
\lc Z_a^{\dag},Z_b^{\dag}\rc+
\lc Z_a,Z_b^{\dag}\rc
\lc Z_a^{\dag},Z_b\rc\bigg)\ , \quad a,b=1,\ldots,4\ . 
\label{zaf}
\ee
The equations of motion of the matrix model and the
constraint take the form
\be
\frac{d^2}{dt^2}Z_a=- \frac{R_{10}^2}{2l_P^6}
\bigg(\lc Z_b^{\dag},\lc Z_b,Z_a\rc\rc
+ \lc Z_b,\lc Z_b^{\dag},Z_a\rc\rc \bigg)\ , 
\label{rat}
\ee
\be
\lc \dot{Z}_a,Z_a^{\dag}\rc+\lc \dot{Z}^{\dag}_a,Z_a\rc=0 \ .
\label{gau}
\ee
A solution with rotation in the planes $Z_a$ can
now be found by analogy with the membrane case, i.e. replacing
$T_{nm}=e^{in\s +im\rho }$ by $J_{nm}$,
 with the matrix ansatz 
\be
Z_a=r_ae^{i\omega_at}J_{\vec{n}_a}\ , \quad a=1,\ldots,d\ ,
\label{azd}
\ee
where $1<d\le 4$.
Consider the case of even $N$. The equations  of motion (\ref{rat})
give the relation
\be
\label{frecs}
\omega_a^2=4\frac{R_{10}^2}{l_P^6} 
\sum_{b=1}^d r_b^2\sin^2\lp\frac{\pi}{N}\lp 
\vec{n}_b\times \vec{n}_a\rp\rp\ ,
\ee
which is the analog of the relation  (\ref{consomega}).
The Gauss constraint (\ref{gau}) is satisfied automatically by the
ansatz (\ref{azd}), using that $J_{\vec m}^\dagger =J_{-\vec m}$.

\subsection{Energy of this configuration}

Let us now evaluate the light-cone energy $P_-$
of the solutions that
we have found above. This is obtained by inserting the
solution (\ref{azd}) into   the Hamiltonian
corresponding to the matrix model (\ref{zaf}). We obtain
\be
P_- =H= \frac{N}{2R_{10}} \sum_{a=1}^d
\omega_a^2r_a^2-\frac{NR_{10}}{l_P^6}
\sum_{a,b=1}^d   r_a^2r_b^2   \sin^2\lp\frac{\pi}{N}\vec{n}_a\times 
\vec{n}_b\rp \ .
\ee
 Using eq. (\ref{frecs}) for the
frequencies, the energy can be
expressed in a  compact form as
\be
P_-=\frac{NR_{10}}{l_P^6}
\sum_{a,b=1}^d r_a^2r_b^2 \sin^2\lp\frac{\pi}{N}\vec{n}_a\times 
\vec{n}_b\rp  \ .
\ee
{}For large $N$, this has the same form as the membrane energy
(\ref{epo}), 
which (using (\ref{consomega}), (\ref{dre})) can be written as
$$
E^2_{\rm mem}={\rm
  const.}\sum_{a,b=1}^dr_a^2r_b^2(l_bk_a-l_a k_b)^2\ .
$$

Let us now consider  explicit cases. In the case of $SU(2)$, the 
simplest non-trivial example, the
generators of the Lie algebra  (\ref{alfar}) coincide with the
usual Pauli matrices, as can be seen in the Appendix.
 For this group we find that any non-trivial configuration
(with non-zero energy) has an energy given by
\be
P_-=\frac{2R_{10}}{l_P^6}
\sum_{a,b=1,a\ne b  }^d r_a^2r_b^2
\sin^2\lp\frac{\pi}{2}\rp =\frac{2R_{10}}{l_P^6}
\sum_{a,b=1,a\ne b}^d r_a^2r_b^2 \ .
\ee 
In a similar way one can obtain the energy for the case of $SU(3)$,
but now it can be seen that there are configurations with different energy.

The matrix model solution can be viewed pictorially as follows.
Recall that the diagonal entries in the matrices $Z_a$
represent strings which begin and end on the same 
D0 brane, whereas the non-diagonal entries represent strings going
from one D0 brane to a different one. For the $SU(3)$ case, there are
three D0 branes. The $Z_a$ can be proportional to any of the eight
generators listed in appendix A. 
Solutions with a  $Z_a$ proportional to a generator with
  three non-diagonal entries represent a  configuration
of three strings joining the three D0 branes, forming a triangle,
which rotate in the plane $Z_a$. In another plane $Z_b$, the
solution can have $Z_b$ proportional to a diagonal generator
$J_{(1,0)}$
or $J_{(2,0)}$,  representing a rotating system of three D0 branes with
strings beginning and ending on the same D0 brane.

\section*{Acknowledgments}

Work 
supported in part by the European Commission RTN programme under 
contract HPNR-CT-2000-00131, by  MCYT FPA 2001-3598, CIRIT GC
2001SGR-00065.
and by the MEC grant AP2002-2415.

 \setcounter{section}{0}
\appendix{Construction of $SU(N)$ generators}

In this appendix we review the construction and relevant 
properties of the $SU(N)$ generators in the
representation that we are using \cite{infinite}. First
 we do this in general and then we give
the explicit expressions for the $SU(3)$ generators. 
These algebra becomes the area-preserving diffeomorphism algebra
of the toroidal membrane in the $N\to\infty$ limit.

A basis for the $SU(N)$ algebras can be built 
from two unitary $N\times N$ matrices,
\be
g\equiv\lp \begin{array}{ccccc} 1 & 0&0&\ldots&0 \\
0 & \omega &0&\ldots&0 \\
0 & 0&\omega^2&\ldots&0 \\
\vdots  & \vdots &\vdots &  &\vdots \\
0 & 0&0&\ldots& \omega^{N-1}  \end{array}\rp,\quad
h\equiv\lp \begin{array}{ccccc} 0 & 1&0&\ldots&0 \\
0 & 0 &1&\ldots&0 \\

\vdots  & \vdots &\vdots &  &\vdots \\
0 & 0&0&\ldots&1 \\
1 & 0&0&\ldots& 0  \end{array}\rp, \quad g^N=h^N=1 \ ,
\ee
where $\omega$ is an $N$'th root of unity with period no smaller than
$N$, that is, $\omega=e^{i4\pi/N}$ for $N$ odd and 
 $\omega=e^{i2\pi/N}$ for $N$ even.
Now using $hg=\omega g h$, it follows that the  unitary $N\times
N$ matrices
\be
J_{\vec{m}}=\omega^{{m_1m_2\over 2}}g^{m_1}h^{m_2}, \quad \vec{m}\equiv
(m_1,m_2)\ ,
\ee
span the algebra of $SU(N)$, that is, they
close under multiplication
\be
J_{\vec{n}}J_{\vec{m}}=\omega^{-{(\vec{n}\times\vec{m})\over 2}}J_{\vec{n}+
\vec{m}} \ ,
\ee
with $\omega$ as defined above, 
so that their commutation rules are given in terms
of trigonometric structure constants,
\be
\label{lieal}
\lc J_{\vec{m}},J_{\vec{n}}\rc=\lp \omega^{\vec{n}\times\vec{m}/2}-
\omega^{-\vec{n}\times\vec{m}/2}\rp J_{\vec{m}+\vec{n}}=-2i\sin\lp
\frac{2\pi}{N}\vec{m}\times\vec{n}\rp J_{\vec{m}+\vec{n}} \ ,
\ee
where the last equality follows for odd $N$ (for even $N$
one has $\pi$ instead of $2\pi$), and the vector product is defined
in the usual way, $\vec{n}\times \vec{m}=n_1m_2-m_1n_2$. 
Another interesting property is that
for the case of three generators whose winding number satisfy
\be
\label{conn}
\vec{n}_1+\vec{n}_2+\vec{n}_3=0 \ ,
\ee
then the algebra of these generators gets simplified and we get a
twisted $SU(2)$ trigonometric algebra
\be
\lc J_{\vec{n}_1}, J_{\vec{n}_2}\rc=-2i\sin\lp
\frac{2\pi}{N}\lp\vec{n_1}\times \vec{n_2} \rp\rp J_{\vec{n}_3}^{\dag}
\ ,
\ee
together with the corresponding permutations. This algebra is twisted
$SU(2)$
in the sense that it is the algebra of $SU(2)$ but with a single
trigonometric
structure constants, since due to the condition (\ref{conn})
it follows that $\vec{n_1}\times \vec{n_2}=\vec{n_1}\times 
\vec{n_3}=\vec{n_2}\times \vec{n_3}$.

{} From their definition, one can check that
all the generators are traceless. Consider, in particular,
 the generators of the form
\be
J_{(m_1,0)}\ , \quad m_1<N
\ee
The trace is given by
\be
\mathrm{tr}\lp J_{m_1,0}\rp=\mathrm{tr}~ g^{m_1}=
\sum_{k=0}^{N-1}\omega^k \ ,
\ee
which indeed vanishes, since $\omega$ is the $N-$th root
of unity.
These generators are normalized as follows,
\be
\mathrm{tr}\lp J_{\vec{n}}J_{\vec{n}}^{\dag}\rp=
\mathrm{tr}\lp J_{\vec{n}}J_{-\vec{n}}\rp=\mathrm{tr}~1=N \ .
\ee

\medskip

Now we compute the explicit expressions for the generators
of the $SU(2)$ and $SU(3)$ algebra in the basis of
 \cite{infinite}. For $SU(2)$ the generators reduce to the
usual Pauli matrices,
\be
J_{(1,0)}=\sigma_3\ ,\quad J_{(0,1)}=\sigma_1\ , \quad
J_{(1,1)}=-\sigma_2\ .
\ee
{}For $SU(3)$ the expressions are more involved, differing from
the
usual canonical basis.
These  generators of the algebra of $SU(3)$  are given
by
\bea
J_{(1,0)}=\lp \begin{array}{ccc} 1 & & \\ & e^{i4\pi/3}& \\ & &
  e^{i2\pi/3}\end{array}\rp, \quad 
J_{(2,0)}=\lp \begin{array}{ccc} 1 & & \\ & e^{i2\pi/3}& \\ & &
  e^{i4\pi/3}\end{array}\rp \nonumber \\
J_{(0,1)}=\lp \begin{array}{ccc}  &1 & \\ & &1 \\1 & &\end{array}\rp, \quad 
J_{(0,2)}=\lp \begin{array}{ccc}  & & 1\\1 & & \\ &1 & \end{array}\rp
\nonumber
\\
J_{(1,1)}=\lp \begin{array}{ccc}  & e^{i2\pi/3} & \\ & &1 \\e^{i4\pi/3} & &
  \end{array}\rp, \quad 
J_{(2,2)}=\lp \begin{array}{ccc}  & &e^{i2\pi/3} \\e^{i4\pi/3} & & \\ & 1&
\end{array}\rp  \\
\\
J_{(2,1)}=\lp \begin{array}{ccc}  & e^{i4\pi/3} & \\ & &1 \\e^{i4\pi/3} & &
  \end{array}\rp, \quad 
J_{(2,2)}=\lp \begin{array}{ccc}  & &e^{i4\pi/3} \\e^{i2\pi/3} & & \\ & 1&
\end{array}\rp \nonumber
\eea
One can explicitly check with this equations the algebra of the
generators, given by
\be
\lc J_{\vec{m}},J_{\vec{n}}\rc=-2i\sin\lp\frac{2\pi}{3}\vec{m}\times
\vec{n}\rp J_{\vec{m}+\vec{n}}
\ee
subject to the periodicity of the generators of the algebra, namely
\be
J_{(m_1+3k_1,m_2+3l_1)}=J_{(m_1,m_2)}
\ee
with $k_a,l_a$ integer numbers.
One can explicitely check that all
these generators are indeed traceless, that is
\be
\mathrm{tr}\lp J_{\vec{n}}\rp=0 \ .
\ee


\end{document}